\documentclass[12pt,a4paper]{article}
\usepackage[margin=2cm]{geometry}
\usepackage[margin=5pt,font=small,labelfont=bf]{caption}
\usepackage{graphicx,amsmath,amssymb,natbib,setspace,
            lineno,xspace,color,subcaption,appendix,authblk}
\definecolor{darkblue}{rgb}{0.0,0.0,0.6}
\usepackage[colorlinks=true,breaklinks=true,linkcolor=darkblue,
            urlcolor=darkblue,anchorcolor=darkblue,citecolor=darkblue]{hyperref}

\newcommand{\aten}{c}
\newcommand{\rxi}{r_\xi}
\newcommand{\sigsat}{\sigma_\text{sat}}
\newcommand{\cmp}{\mathcal{C}}
\newcommand{\smallpar}{\epsilon}
\newcommand{\wvec}{\boldsymbol{k}}
\newcommand{\xvec}{\boldsymbol{x}}
\newcommand{\gvec}{\boldsymbol{g}}
\newcommand{\rhat}{\hat{\boldsymbol{r}}}
\newcommand{\zhat}{\hat{\boldsymbol{z}}}

\newcommand{\strr}{\dot{\varepsilon}}
\newcommand{\Vs}{\boldsymbol{V}}
\newcommand{\vs}{\boldsymbol{v}^S}
\newcommand{\vl}{\boldsymbol{v}^L}
\newcommand{\Div}{\boldsymbol{\nabla}\cdot}
\newcommand{\Grad}{\boldsymbol{\nabla}}
\newcommand{\pdiff}[2]{\dfrac{\partial{#1}}{\partial{#2}}}

\title{Pipe Poiseuille flow of viscously anisotropic, partially molten
  rock}

\author[$\dag$]{Jane Allwright} \author[$\ddag$]{Richard
  F.~Katz\thanks{richard.katz@earth.ox.ac.uk}} \affil[$\dag$]{\small
  Department of Applied Mathematics and Theoretical Physics,
  University of Cambridge, UK} \affil[$\ddag$]{\small Department of
  Earth Sciences, University of Oxford, UK}

\begin{document}
\maketitle

\begin{abstract}
  Laboratory experiments in which synthetic, partially molten rock is
  subjected to forced deformation provide a context for testing
  hypotheses about the dynamics and rheology of the mantle.  Here
  our hypothesis is that the aggregate viscosity of partially molten
  mantle is anisotropic, and that this anisotropy arises from
  deviatoric stresses in the rock matrix. We formulate a model of pipe
  Poiseuille flow based on theory by \cite{takei09a} and
  \cite{takei13}. Pipe Poiseuille is a configuration that is
  accessible to laboratory experimentation but for which there are no
  published results. We analyse the model system through linearised
  analysis and numerical simulations. This analysis predicts two modes
  of melt segregation: migration of melt from the centre of the pipe
  toward the wall and localisation of melt into high-porosity bands
  that emerge near the wall, at a low angle to the shear plane. We
  compare our results to those of \cite{takei13} for plane Poiseuille
  flow; we also describe a new approximation of radially varying
  anisotropy that improves the self-consistency of models over those
  of \cite{takei13}. This study provides a set of baseline,
  quantitative predictions to compare with future laboratory
  experiments on forced pipe Poiseuille flow of partially molten
  mantle.
\end{abstract}

\section{Introduction}

Partially molten regions of the mantle are inaccessible to direct
observations, making it difficult to validate theoretical models for
their mechanics.  Laboratory experiments performed on synthetic mantle
rocks represent a valuable alternative to direct observations.  In
laboratory experiments, when partially molten mantle rocks are
deformed to large strains, bands of high and low volume-fraction of
melt (porosity) emerge spontaneously and remain oriented at a low
angle of $\sim$15--20$^\circ$ to the shear plane
\citep[e.g.][]{king10}.

Modelling this pattern-forming instability is recognised as a means to
validate theoretical models of magma/mantle interaction, regardless of
whether the same instability occurs in Earth's
mantle. \cite{stevenson89} described a one-dimensional model that was
the first to predict the instability under a porosity-weakening
aggregate viscosity; this work preceded and motivated the laboratory
experiments.  Extension to a two-dimensional theory by
\cite{spiegelman03b} predicted porosity band emergence at 45$^\circ$
to the shear plane.  \cite{katz06} obtained theoretical models of
low-angle bands by extending the porosity-weakening viscosity to be
strongly non-Newtonian. However, direct measurements by \cite{king10}
of the stress dependence of the aggregate viscosity in band-forming
experiments were much lower than required by \cite{katz06}, falsifying
their model.  Evidently, although the governing equations permit
formation of high-porosity bands, the details of the pattern depend on
the features of the rheology that is assumed.

In the models noted above, the viscosity of the grain$+$melt aggregate
was assumed isotropic but this need not be the case. \cite{takei09a,
  takei09b} developed a theory for viscous anisotropy of a partially
molten aggregate in which the anisotropy arises from the grain-scale
distribution of melt.  If the aggregate deforms in diffusion creep and
the melt provides a fast pathway for diffusion of grain material, then
a coherent alignment of melt pockets at the micro-scale will give rise
to faster and slower directions for diffusive response to deviatoric
stress at the macro-scale.  According to \cite{takei09a} and
\cite{takei10}, melt-filled pores preferentially align normal to the
direction of largest tensile stress, which reduces the aggregate
viscosity to deformation in the same direction.  On this basis,
\cite{takei09a} and \cite{takei13} formulated a viscosity tensor for
the two-phase aggregate. Analysis of this tensor by \cite{takei09c},
\cite{butler12}, \cite{takei13}, and \cite{katz13} shows that it leads
to a prediction of low-angle porosity bands, consistent with
laboratory experiments.

In laboratory experiments reported by \cite{holtzman03a},
\cite{holtzman07}, \cite{king10}, and \cite{qi13}, the synthetic rocks
subjected to deformation are aggregates of $\sim$10~$\mu$m mantle
olivine and chromite grains, plus 2--5~vol\% basalt or anorthite
powder. The material is raised to a pressure of 300~MPa and
temperature of $\sim$1200$^\circ$C, under which conditions the basalt
or anorthite is molten and resides within the pores between the solid
grains of olivine and chromite.  The samples are held at these
conditions until they reach textural equilibrium, with an
approximately uniform porosity throughout the sample.  They are then
deformed at strain rates of $\sim$10$^{-4}$~sec$^{-1}$ by application
of a deviatoric stress.  The samples are quenched after reaching a
predetermined maximum strain, sectioned, and analysed for the
resulting porosity distribution.  Early experiments by
\cite{holtzman03a} and \cite{holtzman07} were performed in
simple-shear geometry, which imparts an obvious limitation on the
total strain that can be achieved.  Deformation in torsion was
achieved later \citep{king10, qi13} and allows (in theory) for
unlimited amounts of strain.

Both simple shear and torsional deformation were considered in the
theoretical work of \cite{takei13} and \cite{katz13}. Under simple
shear, leading-order flow is lateral and leading-order stress is
initially uniform throughout the experiment.  Under torsional
deformation, the leading-order flow is in the azimuthal direction
around a cylinder; the deviatoric stress is zero at the centre of the
cylinder and largest at the outer radius, giving a gradient directed
radially outward.  \cite{takei13} found that with the inclusion of
anisotropic viscosity, this gradient in deviatoric stress drives melt
migration toward the centre of the cylinder, independent of any
initial porosity variations.  To elucidate this prediction of
``base state segregation,'' they considered a third configuration,
plane Poiseuille flow. In Poiseuille flow, there is gradient in shear
stress from zero at the centre of the flow to a maximum at the outer
edge, where the aggregate abuts the fixed walls.  The geometrical
contrast between plane Poiseuille and torsional deformation enabled
\cite{takei13} to resolve the forces driving melt segregation and make
quantitative predictions of base state segregation.  These predictions
are testable for torsional flow, and indeed early results indicate
agreement with theory \citep{qi13agu, katz13agu}.

Predictions of melt segregation under plane Poiseuille flow are not
readily testable because this configuration is difficult to implement
in the laboratory.  Pipe Poiseuille, however, is an accessible
alternative.  It is therefore the goal of the present manuscript to
apply the formulation of \cite{takei13} for anisotropic viscosity of a
partially molten aggregate to the geometry of pipe Poiseuille
flow. Moreover, the results of such calculations may be relevant to
magma transport within the stem of a mantle plume or a crystal-rich
volcanic conduit, though we do not explore these applications here.
Below we reintroduce the theory and provide new solutions in
cylindrical geometry.  We address the inconsistencies in the analysis
by \cite{takei13} and generate models that are more physically and
mathematically consistent.

The manuscript is organised as follows.  The governing equations are
presented in the next section, followed by a linearised stability
analysis in section~\ref{sec:analysis}.  In
section~\ref{sec:basestate} we consider the leading-order, base state
dynamics for spatially uniform and radially variable anisotropy, and
then compare the results to plane Poiseuille. In
section~\ref{sec:perturbations} we calculate the growth rate of
band-like porosity perturbations. We return to the base state in
section~\ref{sec:numerical} but consider solutions to the fully
nonlinear governing equations.  We discuss our results in light of
previous theoretical and experimental work in
section~\ref{sec:discussion} and provide a summary and conclusions in
\ref{sec:conclude}.

\section{Governing Equations}
\label{sec:governing-equations}

Here we consider a formulation of the equations for coupled
magma/mantle deformation that was presented by \cite{takei13}.  This
formulation differs from other recent versions
\citep[e.g.][]{bercovici01a, rudge11, keller13} in that it allows for
an anisotropic relationship between stress and strain rate.

\subsection{Conservation statements}

The full system of conservation equations consists of two statements
of conservation of mass and two statements of conservation of
momentum.  These are described by \cite{takei13} and can be solved for
the volume fraction of liquid $\phi(\xvec,t)$, solid and liquid
velocity fields $\vs(\xvec,t)$ and $\vl(\xvec,t)$, and liquid pressure
$p^L$.  It is convenient to manipulate the equations to eliminate
$\vl$, resulting in the system
\begin{subequations}
  \label{eq:dimensional_conservation}
  \begin{align}
    \pdiff{\phi}{t} &= \nabla \cdot \left[(1-\phi) \vs\right], \label{eq1}\\
    \Div \vs &= \Div\left[\frac{K}{\eta^L}\left(\Grad p^L - 
      \rho^L \gvec\right)\right], \label{eq2}\\
    \pdiff{p^L}{x_i} &= \pdiff{}{x_j}\left(\sigma_{ij} + p^L
    \delta_{ij}\right)+\overline{\rho} g_i,\label{eq3}
  \end{align}
\end{subequations}
where $K=K(\phi)$ is the permeability, $\eta^L$ is the liquid
viscosity, $\gvec$ is the vector acceleration of gravity,
$\sigma_{ij}$ is the stress tensor of the solid+liquid aggregate
(tension positive), and $\rho^L, \overline{\rho}$ are the liquid and
aggregate densities.  The first equation represents mass conservation
for the solid phase under the assumption of constant solid density;
the second equation is derived from force balance in the liquid phase;
the third equation is force balance for the two-phase aggregate. We
take $\rho^L = \rho^S = \overline{\rho} = \rho = \text{const.}$ in
what follows.  We furthermore take $\eta^L$, and $\gvec$ to be
constant and uniform.

Although we do not solve for the velocity of the liquid explicitly, it
can be obtained directly by substitution of the solution of
\eqref{eq:dimensional_conservation} into the conservation of momentum
equation for the liquid, which is a modified version of Darcy's law
\citep[e.g.,][]{takei13}.

Here we are concerned with the forced flow of the two-phase aggregate
through a cylindrical pipe with a diameter that is much larger than
the grain size of the solid mantle rock.  It is therefore convenient
to write the equations in cylindrical polar coordinates $(r,\psi,z)$,
defined so that $\gvec = -g\zhat$ and the axis of the cylinder lies
along the $z$-axis at $r=0$,
\begin{subequations}
  \label{eq:dimensional_cylindrical}
  \begin{align}
    \pdiff{\phi}{t} &= \frac{1}{r}\pdiff{}{r}\left[r(1-\phi)v_r\right]
    +\pdiff{}{z}\left[(1-\phi)v_z\right] ,\label{eq:dim_cyl_1}\\
    \frac{1}{r}\pdiff{}{r}(rv_r)+\pdiff{v_z}{z}
    &=\frac{1}{r}\pdiff{}{r}\left(r\frac{K}{\eta^L}\pdiff{p^L}{r}\right)
    + \pdiff{}{z}\left[\frac{K}{\eta^L}\left(\pdiff{p^L}{z}+\rho
        g\right)\right], \label{eq:dim_cyl_2}\\
    \pdiff{p^L}{r} &=\pdiff{}{r}\left(\sigma_{rr} +p^L\right) +
    \frac{\sigma_{rr}-\sigma_{\psi\psi}}{r}+\pdiff{}{z}\sigma_{rz} ,
    \label{eq:dim_cyl_3}\\
    \pdiff{p^L}{z} &=\pdiff{}{r}\sigma_{rz} + \frac{\sigma_{rz}}{r} +
    \pdiff{}{z}\left(\sigma_{zz}+p^L\right)-\rho g. \label{eq:dim_cyl_4}
  \end{align}
\end{subequations}
Here we have assumed azimuthal symmetry
($\partial(\cdot)/\partial\psi=0$), zero azimuthal velocity ($v_\psi=0$), and used
$\vs=(v_r,0,v_z)$.  Under this coordinate system and state of
symmetry, the strain-rate tensor for the solid phase is
\begin{equation}
  \strr_{ij}=
  \bordermatrix{~ & r & \psi & z \cr
    r & \pdiff{v_r}{r} & 0 & \dfrac{1}{2}\left(\pdiff{v_r}{z} + 
      \pdiff{v_z}{r}\right) \cr
    \psi & 0 & \dfrac{v_r}{r} & 0 \cr
    z  & \dfrac{1}{2}\left(\pdiff{v_r}{z} + \pdiff{v_z}{r}\right)  & 0
    & \pdiff{v_z}{z} }.
  \label{eq:strainratetensor} 
\end{equation}

\subsection{Constitutive relations}

Closure of the system of partial differential equations
\eqref{eq:dimensional_cylindrical} requires that we specify a
relationship between permeability $K$ and porosity $\phi$, and a
relationship between the bulk stress tensor $\sigma_{ij}$ and the
solid strain-rate tensor $\strr_{ij}$.  For the former we make the
canonical choice appropriate at small porosity,
\begin{equation}
  \label{eq:permeability}
  K(\phi) = K_0\left({\phi}/{\phi_0}\right)^n,
\end{equation}
where $n$ is a constant, usually taken as two or three, and $\phi_0$
is a reference porosity at which the permeability takes its reference
value $K_0$ \citep{mckenzie89, riley91, faul97, wark98}.

Following on the work of \cite{takei13}, we can relate the
second-order stress tensor to the second-order strain-rate tensor via
a fourth-order viscosity tensor,
\begin{equation}
  \label{eq:2}
  \sigma_{ij} +p^L\delta_{ij} = \aten_{ijkl}\strr_{kl},
\end{equation}
where $\delta_{ij}$ is the identity tensor. \cite{takei09a, takei09b}
developed a microstructural model for diffusion creep of a partially
molten rock to relate the macroscopic stress tensor to the viscosity
tensor. They predicted a transversely isotropic viscosity with
rotational symmetry about the axis of maximum tension (the
$\sigma_3$-direction) and anisotropic weakening along this axis; they
further predicted that the magnitude of this anisotropy is a bounded
function of the stress anisotropy, $(\sigma_3-\sigma_1)$.

Consistent with our requirement of azimuthal symmetry of the flow, we
assume that the principal axes of the stress tensor corresponding to
the minimum and maximum tensile stress ($\sigma_1$ and $\sigma_3$,
respectively) lie in the $z$--$r$ plane of the cylindrical coordinate
system for any azimuth $\psi$. On this basis, we specify the
orientation of the anisotropy tensor as the angle $\Theta$ between the
$z$-axis and the direction of maximum tensile stress.  Following
\cite{takei13} in defining $\alpha$ as the magnitude of anisotropy and
requiring $0 \le \alpha \le 2$, we can write the anisotropy tensor as
\begin{align}
  \label{eq:anisotropytensor}
  \aten_{ijkl} = \eta_0\text{e}^{-\lambda(\phi-\phi_0)}\times
  \bordermatrix{~
      & rr & \psi\psi & zz & r\psi & \psi z & zr \cr 
      rr & B & C & A & 0 & 0 & -D \cr 
      \psi\psi & C & C+2 & C & 0 & 0 & 0 \cr 
      zz &A & C & F & 0 & 0 & -E \cr
      r\psi & 0 & 0 & 0 & 1 & 0 & 0 \cr 
      \psi z & 0 & 0 & 0 & 0 & 1 & 0 \cr 
      zr & -D & 0 & -E & 0 & 0 & A-C+1 },
\end{align}
where $\lambda$ is a constant, typically taken in the range 25--30,
and we have defined
\begin{equation}
  \label{eq:atencomponents}
  \begin{aligned}
    A &= \rxi-{2}/{3}-\alpha\cos^2\Theta\sin^2\Theta , \\
    B &= \rxi+{4}/{3}-\alpha\sin^4\Theta ,  \\
    C &= \rxi-{2}/{3} ,  \\
    D &= \alpha\cos\Theta\sin^3\Theta ,  \\
    E &= \alpha\cos^3\Theta\sin\Theta ,  \\
    F &= \rxi+{4}/{3}-\alpha\cos^4\Theta .
  \end{aligned}
\end{equation}
In this set of equations, $\rxi$ is the bulk-to-shear viscosity ratio.
The remaining components of $\aten_{ijkl}$ follow by symmetry of the
tensor.  The tensor $\aten_{ijkl}$ is identical to that used by
\cite{takei13} (their eqn.~4.9) with the relabelling of coordinates
$Z\rightarrow\psi$, $X\rightarrow z$, and $Y\rightarrow r$. It is
important to note that anisotropy ($\alpha>0$) introduces coupling
between shear stresses and normal strain-rates (and, by symmetry,
normal stresses and shear strain-rates).  When $\alpha=0$, the
viscosity tensor reduces to its standard, isotropic form and this
coupling vanishes.

For all calculations in the present manuscript we take $n=3$,
$\phi_0=0.05$, $\lambda=27$, and $\rxi=5/3$. The value for $\rxi$ was
obtained by \cite{takei09a} through microstructural modelling; the
value for $\lambda$ fits experimental data \citep{kelemen97} and is
consistent with the same microstructural model; the value for $n$ is
taken to be $3$ for consistency with previous studies and based on
\cite{wark98} (note, however that the permeability exponent was
recently estimated by \cite{miller14} as $2.6 \pm 0.2$ on the basis of
simulated flow through pore networks obtained by three-dimensional
micro-tomographic scans of texturally equilibrated olivine and basalt.)

\subsection{Scaling and non-dimensionalisation}

Let $H$ be the radius of the cylinder; a typical rate of vertical,
solid-phase flow through the cylinder is then $U = \rho g
H^2/\eta_0$. We use these to introduce rescaled, dimensionless
variables as follows:
\begin{center}
  \begin{tabular}{l l l}
    $\boldsymbol{X} =\dfrac{\boldsymbol{x}}{H}$, & 
    \quad $\Vs =\dfrac{\vs}{U}$, & \quad
    $K^* =\dfrac{K}{K_0}=\left(\dfrac{\phi}{\phi_0}\right)^n$, \\[3mm] 
    $\sigma^*_{ij}=\dfrac{\sigma_{ij}+p^L\delta_{ij}}
    {\left({\eta_0U}/{H}\right)}$, & \quad
    $P =\dfrac{p^L}{\rho gH}$, & \quad
    $\strr^*_{ij} =\dfrac{\strr_{ij}}{\left({U}/{H}\right)}$ ,\\[4mm] 
    $\aten^*_{ijkl} =\dfrac{\aten_{ijkl}}{\eta_0}$, & \quad
    $\tau =\dfrac{t}{\left({H}/{U}\right)}$. & 
  \end{tabular}
\end{center}
We will break with the notation defined above, however, and refer to
the non-dimensional radial and vertical coordinates as $r$ and $z$,
respectively, for the rest of the manuscript.

In writing the non-dimensional equations, it is convenient to define a
ratio of the compaction length, an inherent length scale of
magma/mantle interaction \citep{mckenzie84}, to the system size $H$,
\begin{equation}
  \label{eq:1}
  R = \dfrac{1}{H}\sqrt{\dfrac{(\rxi+\frac{4}{3})\eta_0 K_0}{\eta^L}}.
\end{equation}
Liquid pressure perturbations cause variations in melt flux (and
hence (de)compaction) over a length scale that is less than or equal
to the compaction length \citep{spiegelman93a}.  Hence we expect the
compaction length to influence the scale of emergent features in the
solution.

The governing equations, expressed in terms of non-dimensional
quantities, are
\begin{subequations}
  \label{eq:governing_nondim}
  \begin{align}
    \frac{\partial\phi}{\partial \tau } &=\nabla \cdot \left[(1-\phi)
    \Vs\right], \label{eq:mass}\\
    \nabla \cdot \Vs &= \frac{R^2}{\rxi+{4}/{3}}\nabla
    \cdot\left[K^*(\nabla P+\zhat)\right], \label{eq:cmp} \\
    \frac{\partial P}{\partial r} &=\pdiff{}{r}
      \sigma^*_{rr}+\frac{\sigma^*_{rr}-\sigma^*_{\psi\psi}}{r}+
    \pdiff{}{z}\sigma^*_{rz}, \label{eq:rad}\\
    \frac{\partial P}{\partial z} &=\pdiff{}{r}
    \sigma^*_{rz}+\frac{\sigma^*_{rz}}{r}+\pdiff{}{z}
    \sigma^*_{zz}-1 \label{eq:zed}.
  \end{align}
\end{subequations}
Components of the non-dimensional stress tensor are given by
\begin{equation}
  \label{eq:stresscomponents}
  \sigma_{ij}^* = \text{e}^{-\lambda(\phi-\phi_0)}
  \begin{cases}
    A V_{z,z}+B V_{r,r}+C {V_r}/{r}-D\left(V_{z,r}+V_{r,z}\right) & ij = rr,\\[2mm]
    C V_{z,z} + CV_{r,r}+(C+2)V_r/r & ij = \psi\psi,\\[2mm]
    FV_{z,z}+AV_{r,r}+C{V_r}/{r}-E\left(V_{z,r}+V_{r,z}\right) & ij=zz,\\[2mm]
    -EV_{z,z}-DV_{r,r} +(A-C+1)\left(V_{z,r} +V_{r,z}\right) & ij = rz.
  \end{cases}
\end{equation}
This formulation is valid for radially variable anisotropy parameters
$\alpha$ and $\Theta$, which give rise to radially variable
coefficients $A,B,D,E,F$.

\subsection{Boundary condition}

The pipe wall at $r=1$ is modelled as a no-slip, impermeable, rigid
boundary with $V_r=V_z=0$.  At the centre line $r=0$, we require that
the solution is non-singular, which leads to the symmetry conditions
$V_r = V_{z,r} = P_{,r} = 0$.  We assume an infinitely long pipe, and
hence for the unperturbed base state, we exclude all variation in the
$z$-direction (except for periodic solutions).  Finally, since the
pressure is only constrained up to an additive constant, we choose
that $P=0$ at $r=0$ (without loss of generality).

\section{Analysis}
\label{sec:analysis}

Various authors have employed a linearisation of the governing
equations to study the small-time evolution of porosity that results
from forced deformation. \cite{spiegelman03b} and \cite{katz06}, for
example, considered the stability of plane-wave perturbations to
porosity under a forced, simple-shear flow. \cite{takei09c} and
\cite{butler12} extended this analysis to consider anisotropic
viscosity as formulated by \cite{takei09a}.  This was further extended
by \cite{takei13} to investigate the consequences of anisotropic
viscosity under three flow configurations: simple shear, plane
Poiseuille, and torsion.  Below we compare our results with their
solutions for plane Poiseuille flow.

The strategy for analysis, in all of these studies, is to expand the
solution in a power series of a small parameter $\smallpar\ll1$,
substitute into the governing equations, and balance terms in
$\smallpar^0$ and $\smallpar^1$ separately.  Here we use
\begin{equation}
  \label{eq:expansion}
  \begin{aligned}
    \phi &= \phi_0+\smallpar\phi_1 +O(\smallpar^2), \\
    P &= P_0+\smallpar P_1 +O(\smallpar^2),  \\
    \Vs &= \Vs^{(0)}+\smallpar\Vs^{(1)}+O(\smallpar^2), \\
    \mathcal{C} &= \mathcal{C}_0 +\smallpar\mathcal{C}_1 + O(\smallpar^2),
  \end{aligned}
\end{equation}
where we have defined the compaction rate as $\cmp\equiv\Div\Vs$. The
leading-order terms are called the base state and the first-order terms
are the perturbations.

\subsection{The base state}
\label{sec:basestate}

The base state is initialised with a uniform porosity
$\phi_0$. Under isotropic viscosity ($\alpha=0$), this base state
porosity remains constant with time.  However, under anisotropic
viscosity ($0<\alpha\le2$) and with a spatially varying stress field,
we expect that the base state porosity will evolve in the radial
direction, similar to plane Poiseuille flow \citep{takei13}.  Hence
the base state variables will depend on $r$ and time $\tau$, but will
be independent of $z$.  We seek the instantaneous solution at
$\tau=0$, when the porosity is uniform and the base state permeability
is unity.

The leading-order balances in equations~\eqref{eq:cmp} and
\eqref{eq:rad} can be combined to eliminate the radial pressure
gradient and then integrated to give
\begin{subequations}
  \label{eq:basestate_gov}
  \begin{equation}
    \label{eq:basestate_gov1}
    V^{(0)}_r = \frac{R^2}{\rxi+{4}/{3}}
    \left[\pdiff{}{r}\left(B\pdiff{V^{(0)}_r}{r} + 
        C\frac{V^{(0)}_r}{r} - D\pdiff{V_z^{(0)}}{r}\right)
      + \frac{B-C}{r}\pdiff{V^{(0)}_r}{r} - 2\frac{V^{(0)}_r}{r^2} - 
      \frac{D}{r}\pdiff{V_z^{(0)}}{r}\right];
  \end{equation}
  radial integration of equation \eqref{eq:zed} gives
  \begin{equation}
    \label{eq:basestate_gov2}
    -D\pdiff{V^{(0)}_r}{r}+(A-C+1)\pdiff{V^{(0)}_z}{r} =\frac{r}{2}. 
  \end{equation}
\end{subequations}
Here we have used the boundary conditions and enforced no singularity
at $r=0$.

Given an anisotropy field in terms of $\alpha$ and $\Theta$, equations
\eqref{eq:basestate_gov} can be solved for the base state flow at
$\tau=0$.  We consider two models for the distribution of anisotropy.
The first assumes that anisotropy is uniform \citep{takei13}. The
second model uses a pre-computed, radial variation of $\alpha$ and
$\Theta$ that is based on a hypothesis for the dynamic control of
anisotropy. For the constant-anisotropy case, we compare solutions for
pipe Poiseuille with those for plane Poiseuille flow obtained by
\cite{takei13}.

\subsubsection{Uniform anisotropy}

For an isotropic system, the base state velocity field $\Vs^{(0)}$ is
identical to that of single-phase, incompressible, Stokes flow in the
same geometry.  It is only for non-zero $\alpha$ that the dynamics
lead to divergent solid velocity and hence radial compaction.

The simplest model for the distribution of non-zero anisotropy is
constant $\alpha$ and $\Theta$; this condition was employed by
\cite{takei13}. Uniform anisotropy allows for an analytical solution
to~\eqref{eq:basestate_gov}, which we detail in
appendix~\ref{sec:analytical_solution}.  However, this solution (and
the uniform anisotropy condition itself) violates the expected
symmetry of the problem and leads to a mathematical and physical
singularity. We present examples of the solution nonetheless, for
comparison with previous work and because they are instructive.

\begin{figure}[ht]
  \centering
  \includegraphics[width=12cm]{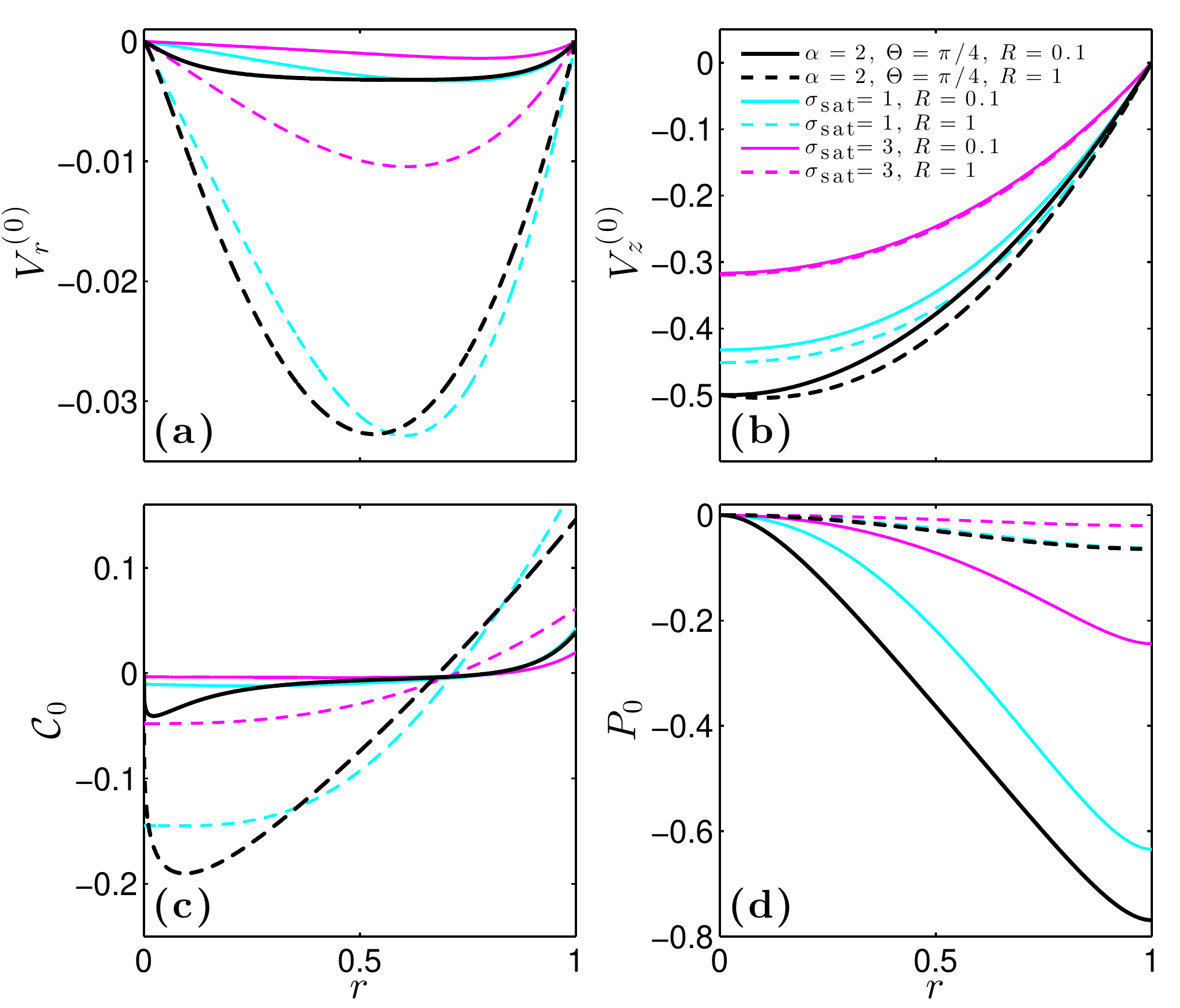}
  \caption{Radial profiles of base state variables for uniform
    anisotropy (black curves; $\alpha=2,\;\Theta=\pi/4$) and
    radially-variable anisotropy (cyan and magenta curves; equations
    \eqref{eq:Theta_r}--\eqref{eq:anconsts}). \textbf{(a)} Radial
    component of the solid velocity $V_r^{(0)}$. \textbf{(b)} Vertical
    component of the solid velocity $V_z^{(0)}$. \textbf{(c)}
    Compaction rate $\cmp_0$. \textbf{(d)} Pressure $P_0$. }
  \label{fig:basestate}
\end{figure}

Black lines in Figure~\ref{fig:basestate} show the base state solution
at $\tau=0$ under uniform anisotropy.  Two representative values of
the dimensionless compaction length are used.  For $R>1$, the
solutions have the same radial structure and saturate at only slightly
larger amplitude than for $R=1$. For $R<0.1$, the boundary layers
narrow and the solution amplitude is reduced. Panel~a shows the radial
component of the velocity; all values are negative, indicating solid
motion toward the centre of the cylinder.  For small compaction
length, there are narrow boundary layers near the centre and wall of
the cylinder, whereas for large compaction length, the radial
component varies throughout the domain.  These features are mirrored
in panel~c, showing the compaction rate.  All curves show that there
is compaction (and hence decreasing porosity) near the centre of the
cylinder and decompaction near the wall. For small compaction length,
there is a range of radii between the compaction boundary layers where
$\cmp_0$ is approximately zero, while for large compaction length,
$\cmp_0$ crosses zero at a point.

Figure~\ref{fig:basestate}c also shows that for constant anisotropy,
$\cmp_0$ has a singular derivative at $r=0$ (black curves).  The
symmetry of the physical problem should lead to the requirement that
$\partial\cmp_0/\partial r=0$ at the centre of the cylinder and, in
fact, that $\cmp_0$ should be an even function of $r$.  
With the constant anisotropy assumption, this
criterion is not met and the radial derivative of the compaction rate
is singular at the centre of the cylinder.

The problem of $\cmp_0$ having a singularity at the centre is also
discernable in both the plane Poiseuille and the torsion analysis of
\cite{takei13}, respectively resulting from the assumptions of
non-zero $\Theta$ and non-zero $\alpha$ at the origin. For torsion and
plane Poiseuille, as for pipe Poiseuille, the horizontal velocity
equation under the uniform anisotropy assumption has a non-smooth
solution \citep[see][]{takei13}. 

Another consideration in evaluating the uniform-anisotropy model is
how well it approximates what it is intended to: the radial variation
in dynamic, stress-induced anisotropy as proposed by
\cite{takei13}. The angle of anisotropy should be defined by the
direction of maximum tensile stress; the magnitude of anisotropy
should approach zero as the magnitude of stress approaches
zero. \cite{takei13} proposed a theory for the dependence of $\alpha$
on the components of the stress tensor, but that theory is
fundamentally nonlinear and not easily incorporated in our analysis.
Therefore, in the next section, we impose $\alpha$ and $\Theta$
\textit{a priori}, as explicit functions of radius.  These functions
are chosen such that they are in approximate agreement with the
retrieved variation of $\alpha$ and $\Theta$, which is computed
\textit{a posteriori} from the base state solution.

\subsubsection{Non-uniform anisotropy}

The model for stress-dependent anisotropy presented in \cite{takei13}
is defined by the following expressions, from which we determine
$\alpha$ and $\Theta$:
\begin{subequations}
  \label{eq:anisotropymodel}
  \begin{align}
    \cos(2\Theta) &= \frac{\sigma_{zz}-\sigma_{rr}}{\sigma_3-\sigma_1},\quad
    \sin(2\Theta) = \frac{2\sigma_{rz}}{\sigma_3-\sigma_1} , \\
    \alpha &= 2\tanh\left(\frac{2(\sigma_3-\sigma_1)}{\sigsat}\right),
  \end{align}
\end{subequations}
where $\sigsat$ is a material parameter and $\sigma_1\le0$ and
$\sigma_3\ge0$ are the values of the principal stresses (tension
positive). Equations~\eqref{eq:anisotropymodel} give
\begin{align}
  \label{eq:alpha_dynamic}
  \alpha &=
  2\tanh\left(\frac{4\sigma_{rz}}{\sigsat\sin(2\Theta)}\right) =
  2\tanh\left(\frac{2(\sigma_{zz}-\sigma_{rr})}{\sigsat\cos(2\Theta)}\right),\\
  \label{eq:Theta_dynamic}
  \Theta &= \frac{1}{2}\arg(2\sigma_{rz}i+\sigma_{zz}-\sigma_{rr}),
\end{align}
in terms of the entries of the stress tensor expressed in system
coordinates.  Here and below, ``$\arg$'' is the argument of the
complex number, and takes values in the range $(-\pi,\pi]$.

\begin{figure}[ht]
  \centering
  \includegraphics[width=15cm]{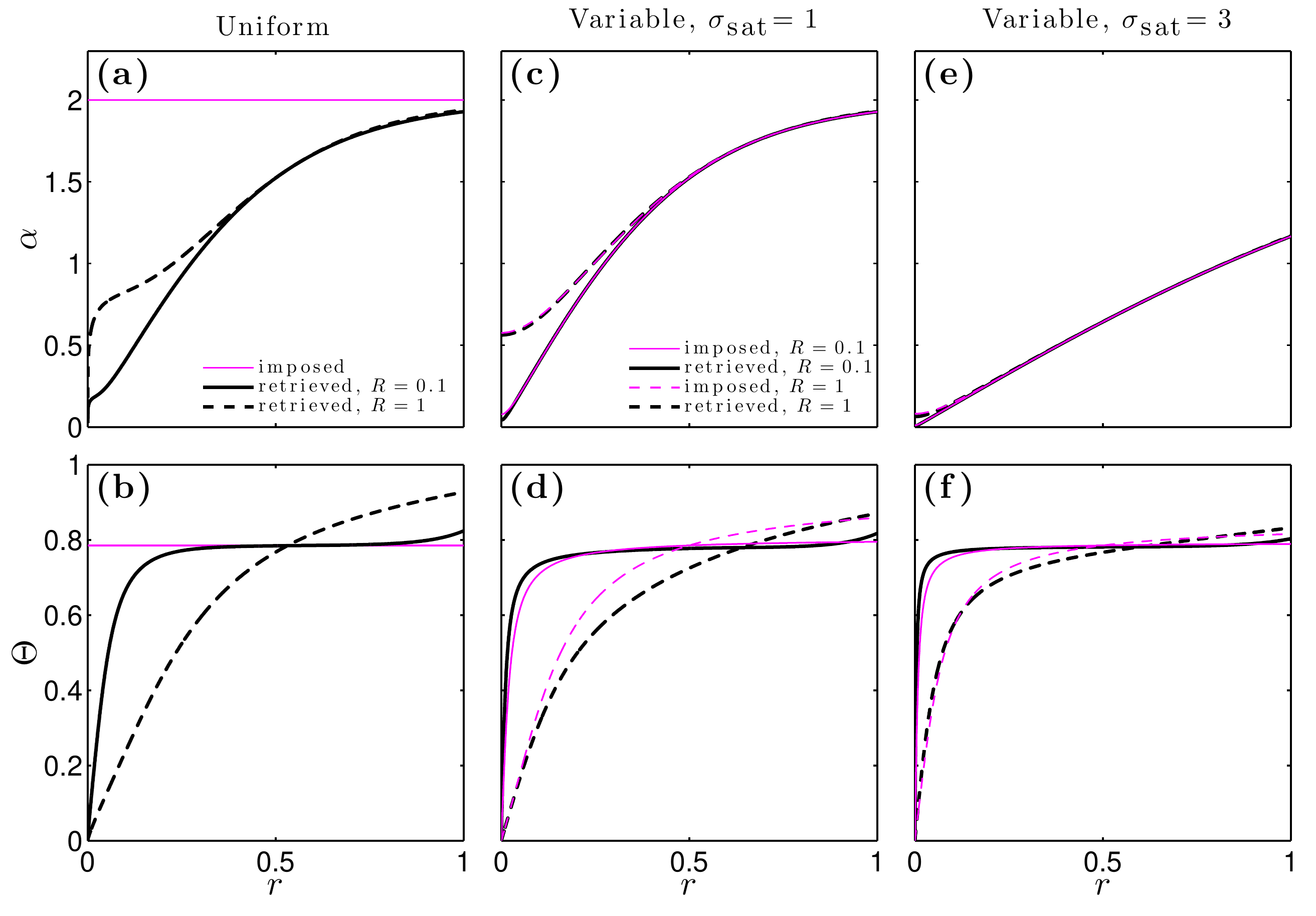}
  \caption{Imposed and retrieved anisotropy parameters $\alpha$ (top
    row) and $\Theta$ (bottom row). Retrieved means computed by
    inserting the radial solution for $\boldsymbol{{V}^{(0)}}$ into
    equations \eqref{eq:alpha_dynamic} and
    \eqref{eq:Theta_dynamic}. The first column has imposed uniform
    anisotropy with $\alpha=2,\;\Theta=\pi/4$; the second and third
    columns have variable anisotropy imposed with equations
    \eqref{eq:Theta_r}--\eqref{eq:anconsts} and with $\sigsat$ of 1
    and 3, respectively.}
  \label{fig:radaniso}
\end{figure}

Figure~\ref{fig:radaniso}a--b show a comparison of imposed (uniform)
anisotropy with the retrieved variation in anisotropy computed by
substituting the stress tensor from the base state solution into
\eqref{eq:alpha_dynamic} and \eqref{eq:Theta_dynamic}.  Focusing
attention on $\Theta$ in panel~b, we see that for the case of $R=0.1$,
there is a region of approximate consistency between the imposed and
retrieved anisotropy, but for $R=1$ there is not.  So we see that
beyond the need for a distribution of anisotropy that respects the
symmetry conditions of the problem, an imposed anisotropy distribution
should be approximately consistent with the consequent base state
distribution of stress.

One way to achieve such consistency is by fixed-point iteration:
imposing the retrieved anisotropy to recompute the flow and then
iterating this process until the difference between anisotropy at
subsequent iterations is below a specified tolerance. However, we are
interested in obtaining approximate analytical forms that may be less
accurate but are of greater utility for subsequent analysis. Below we
propose \textit{a priori} radial forms of $\alpha$ and $\Theta$ to
substitute into the equations, with the aim of achieving the desired
consistency. Specifying the forms of $\alpha$ and $\Theta$ before
solving the differential equations preserves the linearity of the
differential equations in $V^{(0)}_r$ and $V^{(0)}_z$.

A reasonable level of consistency can be achieved with a model of the form
\begin{align}
  \Theta &= \frac{1}{2}\arg
  \left[ri+f_1(r)\right], \quad \alpha =
  2\tanh\left[f_2(r) \right], \label{eq:Theta_r}\\
  \text{with} \quad f_1(r) &=m_1(1-2r) \quad \text{and} \quad f_2(r)
  =\frac{2r}{\sigsat}+m_2 \exp\left(-\frac{2r}{m_2
      \sigsat}\right), \label{eq:anfuncs}
\end{align}
where $m_1$ and $m_2$ are constants that may depend on $R$ and
$\sigsat$.  For large ranges of $R$ and $\sigsat$, suitable
expressions for the constants $m_1$ and $m_2$ are
\begin{subequations}
  \label{eq:anconsts}
  \begin{align}
    \label{eq:em1}
    m_1 &= \tanh R\times\left[0.3\exp\left(-0.6\sigsat\right)+0.03\right],\\
    \label{eq:em2}
    m_2 &= 2m_1/\sigsat .
  \end{align}
\end{subequations}
Note that these expressions for the anisotropy parameters satisfy
$\Theta=0$ and $\partial\alpha/\partial r=0$ at the centre of the
cylinder and we therefore expect the corresponding solution to be more
regular there. However, for an anisotropy model that is completely
smooth at $r=0$ (and hence a completely smooth solution), we would
require $f_1$ and $f_2$ to be even functions of $r$.

Solutions of equations~\eqref{eq:basestate_gov} incorporating radial
variation in anisotropy are obtained numerically, to a tolerance of
$10^{-10}$, using the Chebfun package \citep{driscoll08, chebfunv4,
  trefethen13} for \textsc{Matlab}.

Figure~\ref{fig:basestate}a--b show the components of base state
velocity under an imposed, radially variable anisotropy (coloured
curves).  These differ quantitatively from the uniform anisotropy
case, but the qualitative pattern is unchanged.  The base state
velocity solutions can be used to compute a dynamic anisotropy
(eqns.~\eqref{eq:alpha_dynamic} and \eqref{eq:Theta_dynamic}) to
compare with the imposed anisotropy as a check for self-consistency.

Figure~\ref{fig:radaniso}c--f illustrates the self-consistency of the
anisotropy model defined by equations
\eqref{eq:Theta_r}--\eqref{eq:anconsts}. It shows that for a range of
$R$ and $\sigsat$, the pre-computed anisotropy that is imposed on the
model is approximately consistent with that computed using the
solution.  Also it shows that the models for $\alpha(r)$ and
$\Theta(r)$ satisfy $\partial\alpha/\partial r = \Theta = 0$ at $r=0$.

Figure~\ref{fig:basestate}c shows the base state compaction rate
$\cmp_0$ for uniform and radially variable anisotropy.  For $R=0.1$,
the compaction boundary layer near $r=0$ disappears under variable
anisotropy in favour of a broad, compacting region over most of the
domain. Evidently the solutions with variable anisotropy satisfy the
condition of zero radial derivative of compaction at the centre of the
cylinder. Having achieved both consistency and sufficient regularity,
we conclude that the chosen forms of $f_1$, $f_2$, $m_1$, and $m_2$
are reasonable approximations, and a significant improvement over
uniform anisotropy.

A remaining question about the radially variable model for imposed
anisotropy is how well it agrees with solutions of the full,
nonlinear system of equations with dynamic anisotropy given by
\eqref{eq:alpha_dynamic} and \eqref{eq:Theta_dynamic}. This comparison
is performed below in section~\ref{sec:numerical}, which regards numerical
solutions to the governing equations.  We find excellent agreement at
$\tau=0$ when the numerical model is initialised with uniform
porosity.

\subsubsection{Comparison with the plane Poiseuille base state}
\label{sec:compare_base}

\cite{takei13} obtained an analytical solution for the base state
solid velocity at $\tau=0$ under conditions of plane Poiseuille flow
with uniform anisotropy (constant $\alpha$ and $\Theta=\pi/4$). For a
channel of half-width $H$, the solution can be non-dimensionalised
with the characteristic speed $U=\rho g H^2/\eta_0$ for comparison
with pipe Poiseuille.  This solution is given in terms of
non-dimensional quantities on $0\le x\le 1$ as
\begin{subequations}
  \label{eq:planep}
  \begin{align}
    V^{(0)}_x =&
    \frac{R^2\alpha}{(\rxi+4/3)(4-\alpha)}\left(\frac{\sinh\: [\beta
        x/R]}{\sinh\: [{\beta}/{R}]}+\frac{\sinh\:[{\beta
          (1-x)}/{R}]}{\sinh\:[{\beta}/{R}]}-1\right), \label{eq:planep_vx}\\
    V^{(0)}_z =&
    \frac{2(x^2-1)}{4-\alpha}+\frac{\alpha}{4-\alpha}\left(V^{(0)}_x-(x-1)\left.
        V^{(0)}_{x,x}\right|_{x=0}\right) \label{eq:planep_vz},
  \end{align}
\end{subequations}
where $$\beta =\sqrt{\dfrac{\rxi+\frac{4}{3}}{\rxi+\frac{4}{3} -
    \frac{\alpha}{4-\alpha}}}.$$ 

\begin{figure}[ht]
  \centering
  \includegraphics[width=12cm]{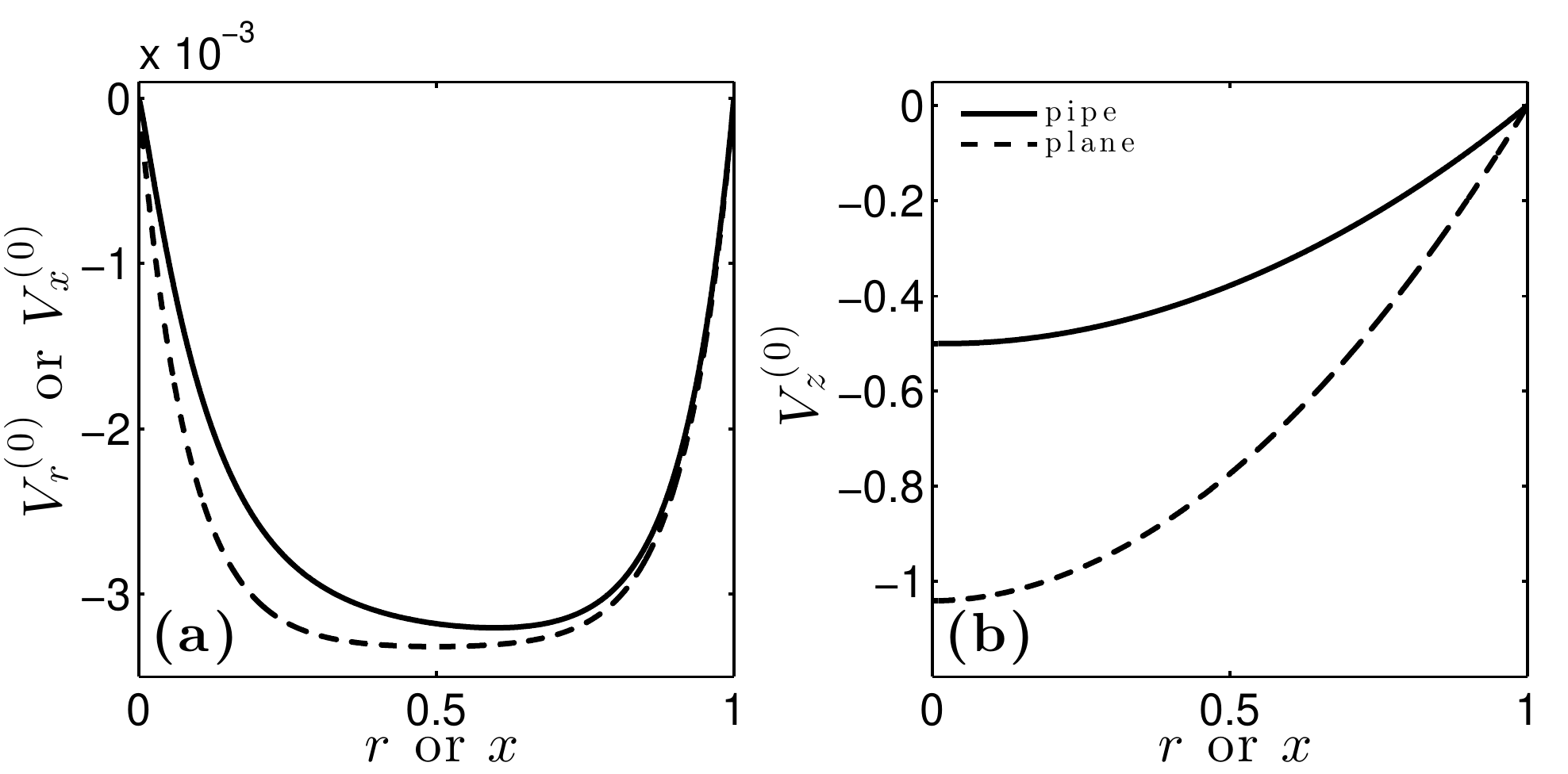}
  \caption{Comparison of base state velocity for pipe and plane
    Poiseuille geometry. Both are computed for uniform anisotropy
    ($\alpha=2,\;\Theta=\pi/4$) and $R=0.1$. \textbf{(a)} Radial or
    lateral component of the velocity.  \textbf{(b)} Vertical
    component.}
  \label{fig:basestatecompare}
\end{figure}

\cite{takei13} assumed constant anisotropy parameters $\alpha$ and
$\Theta$, so for the purpose of comparison, we have used our own
constant anisotropy solution for pipe Poiseuille, with the identical
parameter values. Therefore both the planar and cylindrical solutions
have physical inconsistencies at the centre, but their comparison
nevertheless demonstates a broad similarity between the flows, and
suggests how they scale relative to each other.

Figure~\ref{fig:basestatecompare} shows profiles of solid velocity
components for plane Poiseuille flow, plotted alongside profiles for
pipe Poiseuille flow with uniform anisotropy. The vertical velocity
components have the same shape, but differ by a factor of two in
magnitude; this is what we would expect from the analytical solutions
to single phase, isoviscous Poiseuille flow in the two flow
geometries. The horizontal components are also similar in shape, and
as $R$ increases the horizontal flow in cylindrical geometry becomes
smaller relative to that in planar geometry. However, for $R\lesssim
0.1$, the horizontal velocities are approximately equal in magnitude,
as shown in figure~\ref{fig:basestatecompare}a. In other words, for
these smaller values of $R$, if we normalise by the magnitude of the
vertical flow, we predict stronger lateral flow in the cylindrical
geometry.

\subsection{Growth of porosity perturbations}
\label{sec:perturbations}

Having obtained solutions for the base state variables, we now turn
our attention to the terms of order $\smallpar$ in equations
\eqref{eq:expansion}. These represent perturbations to the base state;
we will analyse them by seeking harmonic solutions that can grow or
decay exponentially with time. There is no universally accepted method
for analysing the linear stability of a time-dependent base state
\citep{doumenc10}.  For simplicity, we consider perturbation growth
only at $\tau=0$, and therefore take $\phi_0$ to be constant and uniform
in solving for the evolution of perturbations.  The calculations in
this section are valid for uniform and radially variable anisotropy.

Equations to constrain the $O(\smallpar)$ terms are obtained by
substituting the expansion \eqref{eq:expansion} into the governing
equations \eqref{eq:governing_nondim}.  The leading-order terms
already balance and the terms of $O(\smallpar^2)$ can be neglected,
leaving equations for $\partial\phi_1/\partial\tau$, $P_1$, and
$\boldsymbol{V}^{(1)}$; these equations are given in
Appendix~\ref{sec:perturbation_growth}. We consider porosity
perturbations of the form
\begin{equation}
  \label{eq:porosityperturb}
  \phi_1(r,z,\tau) = \exp\left[i\wvec\cdot\left(\xvec
      -\int\limits_0^\tau {\boldsymbol{V}^{(0)}}
      \mathrm{d}t\right)+\Psi(r,z,\tau)\right],
\end{equation}
where the wave-vector is $\wvec = k_r\rhat + k_z\zhat$ with $k_r$ and
$k_z$ constants.  Equation~\eqref{eq:porosityperturb} represents cones
of locally harmonic waves moving passively in the base state flow
$\boldsymbol{V}^{(0)}$ with a time-dependent log-amplitude $\Psi$; the
tips of the cones are located at $r=0$ and point upward for
$k_r,k_z>0$. We define
\begin{align}
  k &\equiv \sqrt{k_r^2+k_z^2} , \label{eq:mag_wv}\\
  s+i\Omega &\equiv \frac{\partial\Psi}{\partial\tau} ,  \label{eq:growth_rate}\\
  \text{and we require} \quad \nabla\Psi \text{, } \nabla (s+i\Omega) &= o(k)
  \text{ as } k\rightarrow\infty, \label{eq:slowvary}
\end{align}
where a quantity $q$ that is $o(k)$ satisfies $q/k\rightarrow 0$ as
$k\rightarrow\infty$. The last equation states that both $\Psi$ and
the growth rate of porosity perturbations $s+i\Omega$ are slowly
varying in space, in the sense that they vary on a length scale that
is much longer than the wavelength of perturbations. Because the
governing system is linear at each order of $\smallpar$, we can relate
other variables to $\phi_1$ as
\begin{equation}
  \boldsymbol{V}^{(1)}=\tilde{\boldsymbol{V}}\phi_1 \text{, }\quad
  P_1=\tilde{P}\phi_1 \label{eq:othervars},
\end{equation}
with $\tilde{V_r}$, $\tilde{V_z}$, and $\tilde{P}$ also only slowly
varying. In the following analysis we do not perturb the quantities
$A$ to $F$; the analysis is therefore suitable for $\alpha$ and
$\Theta$ either constant or specified \textit{a priori}, but not
dynamically variable.

Fixing $\tau=0$, taking $k\rightarrow\infty$, and neglecting
all but terms of leading order in $k$, the equations can be inverted
to give expressions for $\tilde{V_r}$, $\tilde{V_z}$, and $\tilde{P}$
that are valid to leading order in $k$.  Then, to obtain the growth
rate, we use \eqref{eq:mass} at $O(\smallpar)$ leading to
\begin{equation}
  s+i\Omega\sim (1-\phi_0)\left(ik_r\tilde{V_r}+ik_z \tilde{V_z}\right)
  -\nabla\cdot\boldsymbol{V}^{(0)} \label{eq:growthrate_equation}
\end{equation}
as $k\rightarrow\infty$ (here and below we use the symbol $\sim$
to mean ``is asymptotic to.'') Substituting for $ik_r\tilde{V_r}+ik_z
\tilde{V_z}$ we find that at $\tau=0$ and to leading order in $k$, the
growth rate is
\begin{equation}
  \label{eq:growthrate}
  s\sim \frac{1-\phi_0}{N_3N_5-{N_4}^2}
  [(N_1N_5-N_2N_4)W_2+(-N_1N_4+N_2N_3)W_3] -\cmp_0 
\end{equation}
where
\begin{equation}
  \label{eq:growthrate_factors}
  \begin{aligned}
    N_1 &= ik_r , \\
    N_2 &=ik_z  , \\
    N_3 &= Bk_r^2+(A-C+1)k_z^2-2Dk_rk_z, \\
    N_4 &= -Dk_r^2-Ek_z^2+(2A-C+1)k_rk_z , \\
    N_5 &= (A-C+1)k_r^2+Fk_z^2-2Ek_rk_z, \\
    W_2 &= -\lambda ik_r\left(BV^{(0)}_{r,r}
    + C{V^{(0)}_r}/{r} - DV^{(0)}_{z,r}
      \right)-\lambda ik_z\left[-DV^{(0)}_{r,r}+(A-C+1)V^{(0)}_{z,r}\right], \\
    W_3 &= -\lambda ik_r\left[-DV^{(0)}_{r,r} + (A-C+1)V^{(0)}_{z,r}\right] 
    - \lambda ik_z\left(AV^{(0)}_{r,r}+C{V^{(0)}_r}/{r}-EV^{(0)}_{z,r}\right).
  \end{aligned}
\end{equation}
In obtaining this solution we find that $\Omega$ is $o(1)$ as
$k\rightarrow\infty$ and, since $s$ is $O(1)$, we have found all the
terms of $s+i\Omega$ that do not decay in the $k\rightarrow\infty$
limit of short wavelength. More details of the above calculations are
provided in Appendix~\ref{sec:perturbation_growth}.

\begin{figure}[ht]
  \centering
  \includegraphics[width=\textwidth]{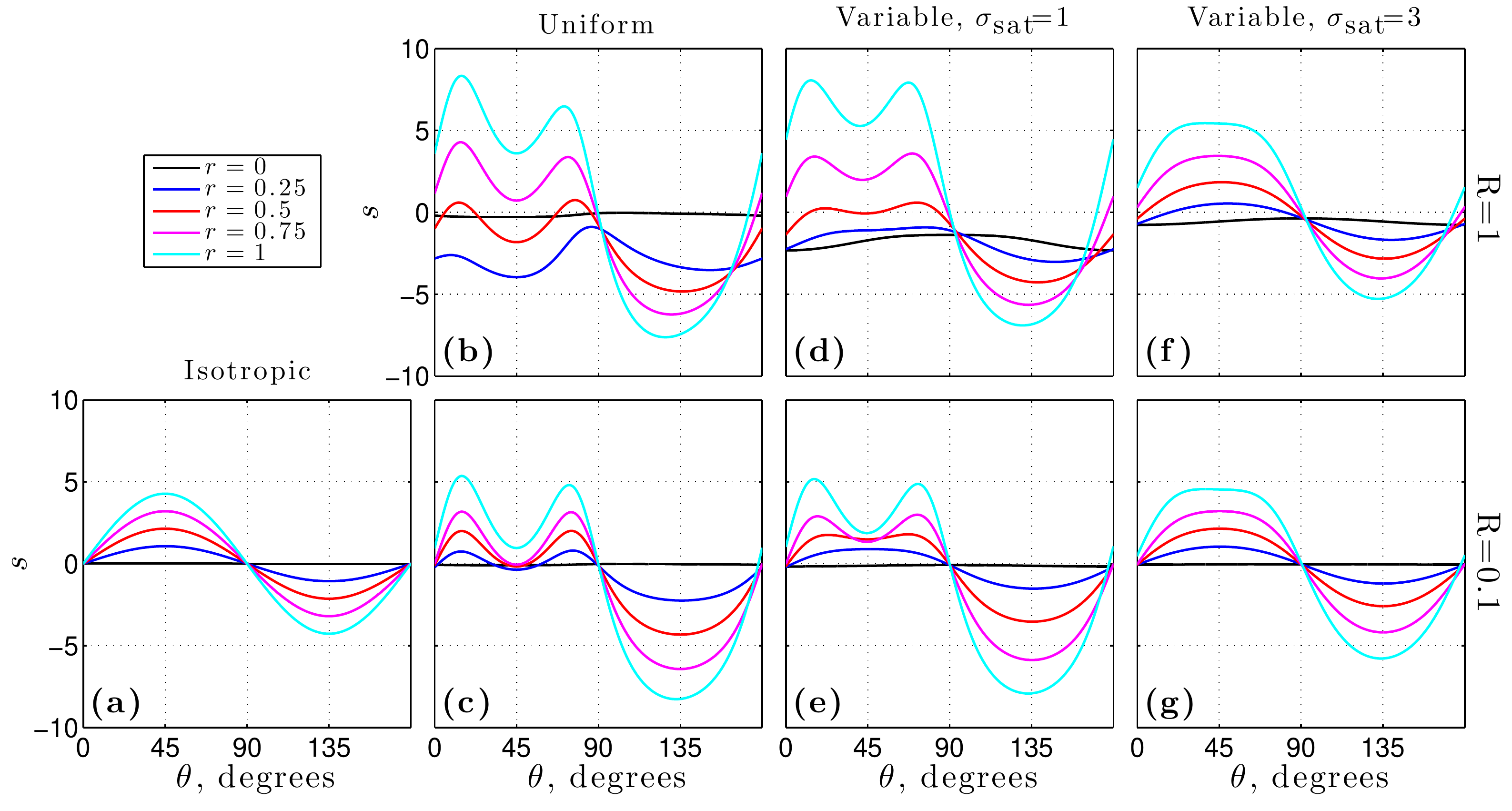}
  \caption{Growth rate $s$ of perturbations as a function of
    perturbation angle $\theta$ from
    equation~\eqref{eq:def_theta}. Columns (labelled above) represent
    different anisotropy scenarios; rows (labelled at right) represent
    different values of non-dimensional compaction length
    $R$. \textbf{(a)} $\alpha=0$; \textbf{(b)--(c)} $\alpha=2,\,
   \Theta=\pi/4$; \textbf{(d)--(e)} $\sigsat=1$; \textbf{(f)--(g)}
    $\sigsat=3$. Panels (d)--(g) use radially variable anisotropy
    $\alpha(r),\Theta(r)$ computed with
    equations~\eqref{eq:Theta_r}. All curves are computed with
    $\lambda=27,\,\phi_0=0.05$.}
  \label{fig:perturbations}
\end{figure}

In the case of isotropic viscosity, the above result
\eqref{eq:growthrate} for the growth rate $s$ reduces to
\begin{equation}
  \label{eq:growthrate_isotropic}
  s\sim\frac{(1-\phi_0)\lambda r k_r k_z}{(\rxi+{4}/{3})k^2}
\end{equation}
as $k\rightarrow\infty$.  Let us write
\begin{equation} 
  \label{eq:def_theta}
  \wvec=k(\cos\theta\rhat + \sin\theta\zhat) ;
\end{equation}
then as $k\rightarrow\infty$, the growth rate $s$ is proportional to
$\sin(2\theta)$ and so is largest for the perturbations with angle
$\theta=45^\circ$.  This is evident in
Figure~\ref{fig:perturbations}a. The equation for $s(\theta)$ under
isotropic viscosity is identical to that obtained for plane
Poiseuille flow by \cite{takei13} (up to a scaling constant and
replacing $r$ with $Y$).

Figure~\ref{fig:perturbations}b--c show the growth rate $s$ of
perturbations at angle $\theta$ for constant, non-zero anisotropy,
given by values of $\Theta={\pi}/{4}$ and $\alpha=2$. Clearly
$45^\circ$ is no longer the most favourable angle for growth. Focusing
attention on the cyan curves representing band growth at the outer
radius of the cylinder, we see that one effect of the anisotropy is to
split the single growth rate peak of panel~a into two peaks
(corresponding to the fastest growing disturbances at $\tau=0$), one
at an angle less than $45^{\circ}$ and one at an angle greater than
$45^{\circ}$. This effect was also found by \cite{takei13} for plane
Poiseuille.  \cite{takei13} showed that the positions and relative
heights of the two peaks depend on the value of the anisotropy angle
$\Theta$. In the case $\Theta={\pi}/{4}$, the peaks occur at angles of
approximately $15^{\circ}$ and $75^\circ$.  For values of constant
$\Theta$ less than ${\pi}/{4}$, the low-angle peak is dominant, and
occurs at an angle larger than $15^{\circ}$.

The uniform anisotropy calculations in
Figure~\ref{fig:perturbations}b--c also show a clear trend with
radius.  Growth rates are fastest for low-angle porosity bands located
at the outer radius because the strain rate is largest there.  The
shear strain rate goes to zero at the centre of the cylinder and hence
we expect $s\approx 0$ there (the contributions by $\cmp_0$ and
$V^{(0)}_{r,r}$ to $s$ are small but non-zero at $r=0$). However, for
antithetical porosity bands ($\theta<90^\circ$) at slightly larger
radii (e.g.,~$r=0.25$), we see that the growth rate can be negative,
meaning that perturbations decay.  This is due to the contribution of
base state compaction $\cmp_0$ in eqn.~\eqref{eq:growthrate}. This
effect is stronger for $R=1$ because the base state compaction is
larger in amplitude (Fig.~\ref{fig:basestate}c). For non-dimensional
compaction lengths greater than unity, the base state compaction rate
saturates in amplitude \citep{takei13}.

These same effects are evident in panels d--g of
Figure~\ref{fig:perturbations}, where the anisotropy varies radially
according to equations~\eqref{eq:anfuncs}. For $\sigsat=1$, $\alpha$
reaches saturation at the outside of the cylinder, giving large growth
rates. In contrast, for $\sigsat=3$, anisotropy is relatively muted
and hence growth rates are overall smaller and the two peaks merge
into a single, broad peak growth rate.  The shift from small $\Theta$
at the centre of the cylinder to $\Theta\gtrsim\pi/4$ at the wall can
be discerned in panel~d, where the low-angle peak is shifted to larger
$\theta$ at $r=0.25$ and smaller $\theta$ at $r=1$.

The general systematics of perturbation growth rates are consistent
for constant and radially variable anisotropy, as well as for plane
and pipe Poiseuille flow.  Indeed when we compare equations for the
growth rates in plane and pipe flow, we see that differential
operators of $\partial_x$ for plane flow become $\partial_r + r^{-1}$
in pipe flow.  When these operators are applied to the perturbation
variables in the limit of $k\rightarrow\infty$, the extra term in
$r^{-1}$ (coming from the curved geometry) is neglected because it is
of a lower order in $k$ than the radial derivative.  Nevertheless,
there are cylindrical terms (i.e.,~$V_r^{(0)}/r$) that appear in $W_1$
and $W_2$ (see eqn.~\eqref{eq:growthrate_factors}), showing that $s$ does
depend in some way on the geometry of the flow.

In fact, the most important difference in perturbation growth between
pipe and plane Poiseuille comes from the overall scaling of the
flows. In this manuscript we scale velocity with $\rho g H^2/\eta_0$
whereas \cite{takei13} use a value twice as large, $2\rho g
H^2/\eta_0$, reflecting the stronger vertical flow in plane geometry
(Fig.~\ref{fig:basestatecompare}b, above). This means that our $\tau$
is half that of \cite{takei13}. Therefore, although the nondimensional
growth rate of perturbations in Figure~\ref{fig:perturbations} is
approximately equal to that in Figure~9 of \cite{takei13}, the
\textit{dimensional} growth rate of perturbations in pipe geometry is
about half that of plane geometry, if the pipe and channel have equal
diameter and thickness, respectively. This difference arises from the
larger vertical shear associated with plane Poiseuille flow
(Fig.~\ref{fig:basestatecompare}b).

\begin{figure}[ht]
  \centering
  \includegraphics[width=14cm]{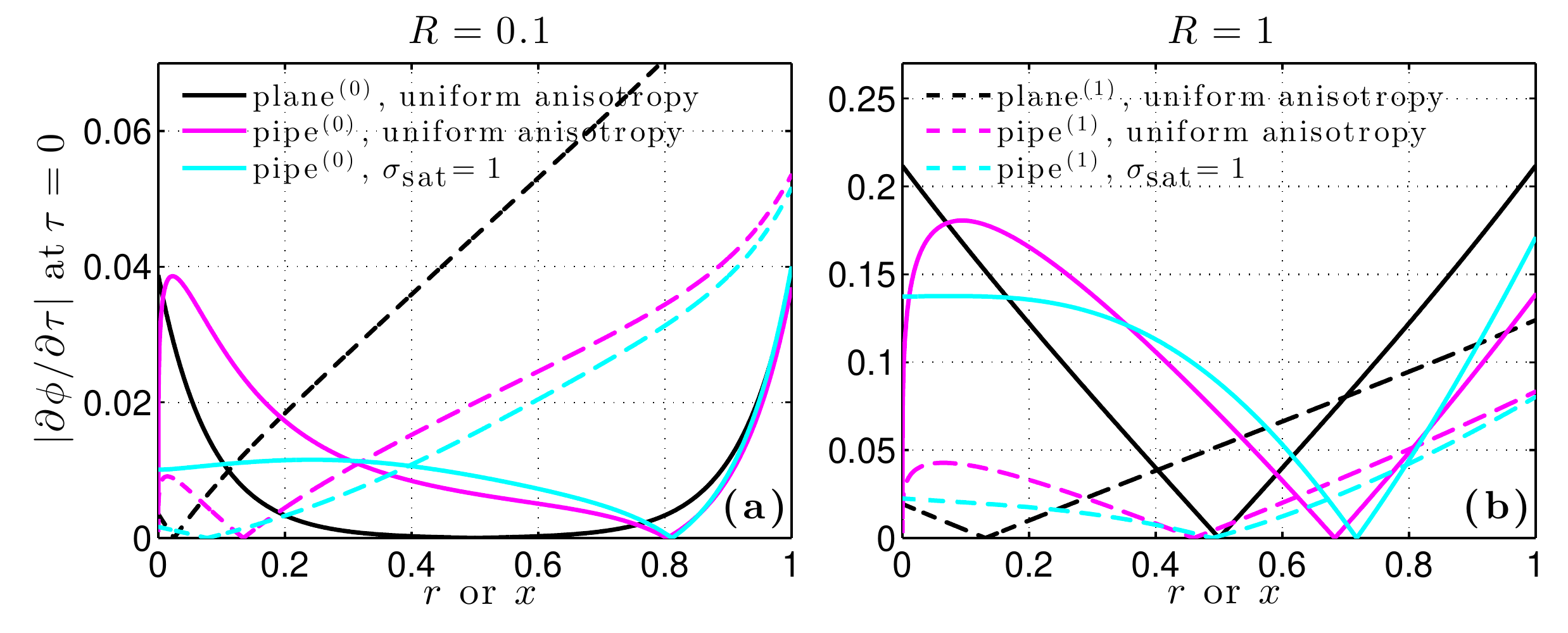}
  \caption{The magnitude of terms in the porosity evolution equation
    \eqref{eq:porosityevolutionequation}. Solid lines show the
    magnitude of base state segregation $\vert(1-\phi_0)\cmp_0\vert$;
    dashed lines show the magnitude of the perturbation growth rate
    $\vert\smallpar s\phi_1\vert$. Plane Poiseuille with uniform
    anisotropy is plotted in black as a function of $x$ (scaled with
    $U=\rho g H^2/\eta_0$ and computed with $\Vs^{(0)}$ from
    \eqref{eq:planep} above). Pipe Poiseuille is plotted in magenta
    and cyan as a function of $r$, computed with uniform and variable
    anisotropy, respectively. Pipe and plane flow are computed for a
    pipe and channel of equal width. The perturbation angle is chosen
    as $\theta=15^\circ$. For uniform anisotropy we use $\alpha=2$ and
   $\Theta=\pi/4$ while for variable anisotropy we use $\sigsat=1$.
    Other parameter values are $\phi_0=0.05$ and
    $\smallpar=0.2\phi_0$. Panel \textbf{(a)} has $R=0.1$; panel
    \textbf{(b)} has $R=1$. }
  \label{fig:porosityevolution}
\end{figure}

A more detailed comparison of the rates of base state segregation and
perturbation growth is given in Figure~\ref{fig:porosityevolution}.
This figure shows the magnitude of terms on the right hand side of the
porosity evolution equation
\begin{equation}
  \label{eq:porosityevolutionequation}
  \pdiff{\phi}{\tau} = \left(1-\phi_0\right)\cmp_0 
  +\smallpar s \phi_1.
\end{equation}
We expect that the local behaviour of the model is predicted by the
term with the larger magnitude. Following \cite{takei13}, we take
$\smallpar=0.2\phi_0$ and consider a fixed perturbation angle
$\theta=15^\circ$ --- this being an optimum value of the growth rate
(Fig.~\ref{fig:perturbations}). The growth rate under plane Poiseuille
flow was obtained by \cite{takei13} but here it is scaled by $U/H=\rho
g H/\eta_0$ (as for pipe flow). The figure predicts that in general,
porosity bands are expected to be prominent for small compaction
length (panel a) while at larger compaction lengths, base state
segregation dominates (panel b).

We can also compare the two modes of segregation for pipe and plane
flow. For $R=0.1$ and adjacent to the no-slip wall, pipe and plane
flow have approximately equal rates of porosity change due to base
state segregation; but through much of the domain, pipe flow has more
rapid porosity change by base state segregation.  This is in contrast
to the rate due to perturbation growth, which is greater for plane
Poiseuille throughout the domain. Hence we expect that for small
compaction length, high porosity bands are less prominent under pipe
Poiseuille flow than under plane Poiseuille.  For $R=0.1$ a similar
comparison holds, though it is muted: the ratio of magnitude of the
two terms on the right-hand side of
eqn.~\eqref{eq:porosityevolutionequation} is approximately the same
for pipe and plane flow.

\section{Solutions of the full, nonlinear equations}
\label{sec:numerical}

\begin{figure}[ht]
  \centering
  \includegraphics[width=12cm]{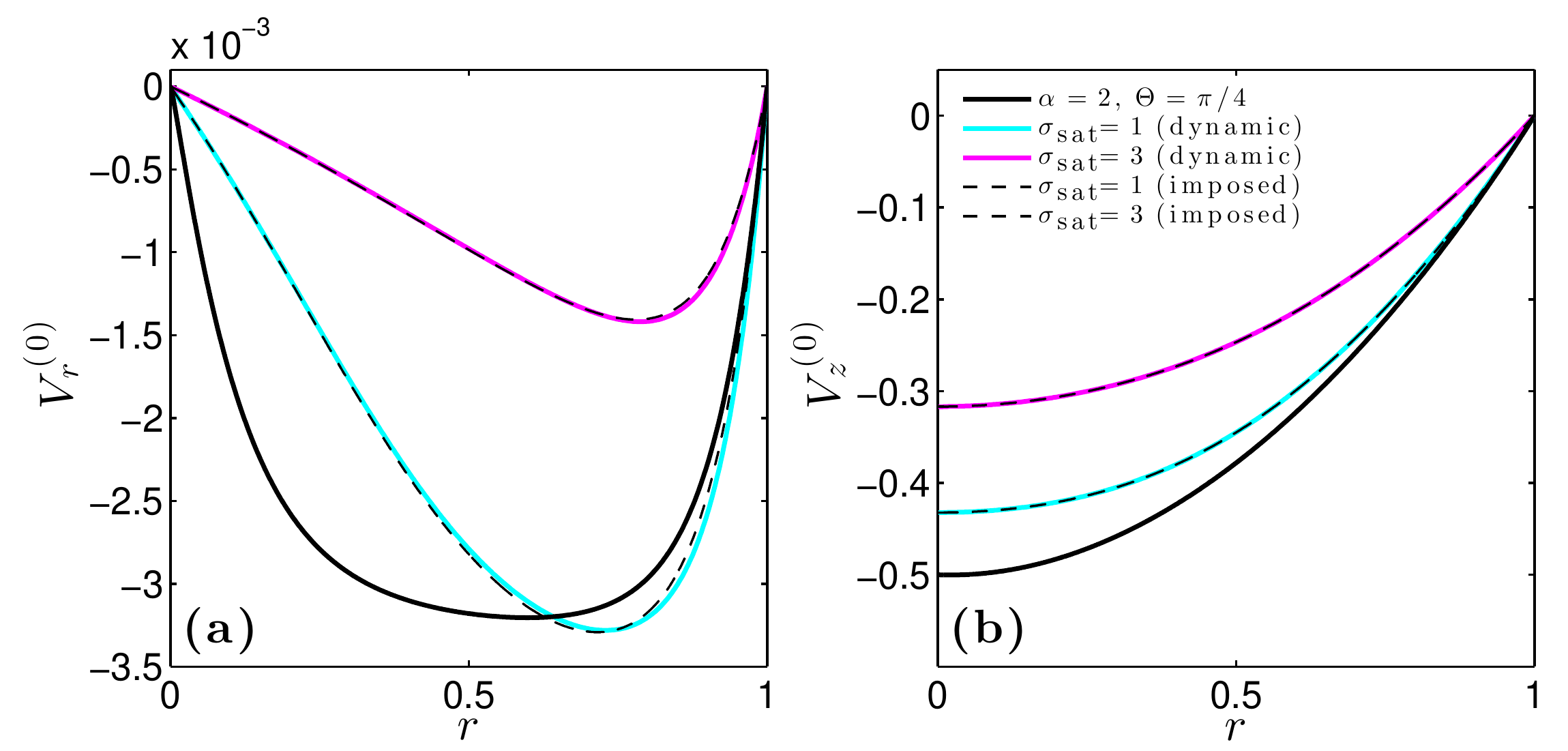}
  \caption{Comparison of radial profiles obtained by solution of the
    full, nonlinear system with dynamic anisotropy (coloured curves,
    eqns.~\eqref{eq:governing_nondim} \&
    \eqref{eq:alpha_dynamic}--\eqref{eq:Theta_dynamic}) to profiles
    from the base state model with imposed anisotropy (black curves,
    eqns.~\eqref{eq:basestate_gov} \&
    \eqref{eq:anfuncs}). \textbf{(a)} Radial component of
    velocity. \textbf{(b)} Vertical component.  All calculations use
   $R=0.1,\,\lambda=27,\,\phi_0=0.05,\,n=3$; the fully nonlinear
    solutions have 400 grid-cells in the radial direction.}
  \label{fig:numericalcompare}
\end{figure}

For the solutions considered in previous sections, at $\tau=0$ we impose the
anisotropy \textit{a priori} to keep the equations linear. However,
equations~\eqref{eq:alpha_dynamic} and \eqref{eq:Theta_dynamic}
provide a recipe for computing the \textit{dynamic} anisotropy---the
pointwise values of $\alpha$ and $\Theta$ that are in equilibrium with
the instantaneous stress tensor of the aggregate. This formulation of
the viscosity is nonlinear and hence we abandon the linearised
governing equations and return to the full, nonlinear system
\eqref{eq:governing_nondim}.  We proceed by discretising the
governing equations with a finite volume approximation and solving the
resulting system of nonlinear algebraic equations using algorithms
provided by the Portable, Extensible Toolkit for Scientific
Computation \citep[PETSc,][]{petsc-homepage, petsc-manual, katz07}.
Details are provided in Appendix~\ref{sec:numerical_implementation}.

Although our discretisation and code implementation allow for a
two-dimensional ($r$--$z$) domain, we consider only one-dimensional
profiles to focus attention on the nonlinear evolution of the base
state.  The band-forming instability is avoided by considering an
initial condition of porosity that is spatially uniform to machine
precision. Simulations that are initiated with a small amount of
white-noise variation added to the background porosity do produce high
porosity bands.  As expected from the linearised theory, they are at
low angle to the shear direction, appear close to the pipe wall, and
are of smaller amplitude than those in plane Poiseuille
\citep{katz13}.  These two-dimensional solutions are not reproduced
here.

Figure~\ref{fig:numericalcompare} compares one-dimensional solutions
to the nonlinear governing equations at $t=0$ with solutions computed
using the leading-order equations \eqref{eq:basestate_gov} and imposed
anisotropy for $R=0.1$ and $\sigsat=1,3$. The excellent match between
calculations with dynamic and imposed anisotropy indicates that the
numerical solution is accurate (small differences are the result of
imperfection in the imposed anisotropy model in
equations~\eqref{eq:Theta_r}--\eqref{eq:anconsts} with respect to the
dynamic determination of self-consistent anisotropy).

\begin{figure}[ht]
  \centering
  \includegraphics[width=\textwidth]{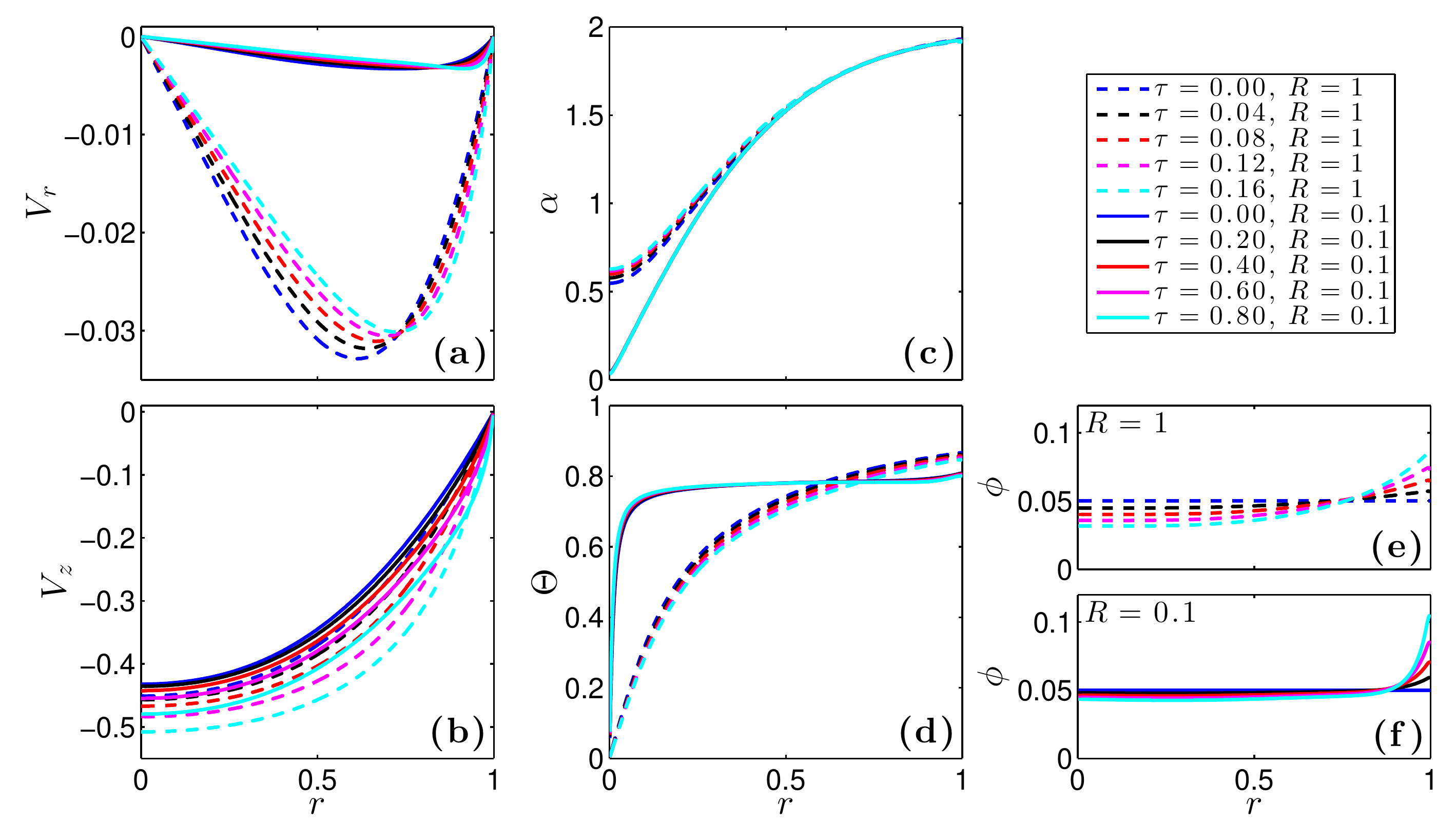}
  \caption{Solutions to the full governing equations at various times
    $\tau$. \textbf{(a)} Radial component of velocity. \textbf{(b)}
    Vertical component. \textbf{(c)} Anisotropy
    magnitude. \textbf{(d)} Anisotropy angle. \textbf{(e)} Porosity
    for $R=1$. \textbf{(f)} Porosity for $R=0.1$. All calculations use
    $\lambda=27,\,\phi_0=0.05,\,n=3$; the fully nonlinear solutions
    have 400 grid-cells in the radial direction.}
  \label{fig:evolution}
\end{figure}

Figure~\ref{fig:evolution} shows the evolution of solutions for
$R=0.1$ (solid curves) and $R=1$ (dashed curves) over a finite time
interval. The time interval is longer for $R=0.1$ because the radial
segregation rate is slower (e.g.~Fig.~\ref{fig:basestate}c).  For both
values of the non-dimensional compaction length, however, we see that
porosity and shear localise toward the no-slip wall.  This was also
the case for plane Poiseuille flow \citep{takei13}.  If the
simulations are allowed to evolve forward beyond the time interval
shown, the porosity continues to localise toward the wall, reducing
the aggregate viscosity there.  Shear is therefore focused at the wall
while strain rates in the interior of the flow decrease.  The system
rapidly reaches a plug-flow configuration where all deformation is
located in a narrow zone of high porosity along the wall. It should be
noted, however, that the high porosities reached in this scenario
violate assumptions used to derive the governing equations (i.e., that
the solid forms a contiguous matrix and that shear stresses in the
liquid phase are negligible).

\section{Discussion}
\label{sec:discussion}

Pipe Poiseuille flow of a two-phase aggregate with anisotropic
viscosity is related to torsional and plane Poiseuille flow, but it
differs in important ways.  It shares a cylindrical geometry with
torsional flow, including base state, compressional hoop stress
($\sigma_{\psi\psi}<0$).  In the case of torsional flow
\citep{takei13}, the compressional hoop stress is caused by viscous
anisotropy in the tangential ($\psi$--$z$) plane. Both the $\sigma_1$
and $\sigma_3$ directions lie within this plane (to leading order), in
an arrangement that is identical to that of simple shear. The
$\sigma_1$ (compressional) stress is associated with a large viscosity
while the $\sigma_3$ (tensile) stress is associated with a reduced
viscosity.  Hence the imposed shear results in a net compressive
stress in the tangential plane: a negative hoop stress.  This causes a
positive radial pressure gradient (eqn.~\eqref{eq:rad} above) driving
liquid inward (and solid outward).

In contrast, under pipe Poiseuille, the maximum compressive and
tensile stresses lie in the $z$--$r$ plane and are the result of the
gravitational body force (last term in eqn.~\eqref{eq:zed} above).
These stresses increase in magnitude with $r$. Combined with a tensile
viscosity that decreases radially with increasing deviatoric stress,
this results in a negative radial pressure gradient.  The pressure
gradient, in turn, drives liquid outward toward the pipe wall (and
solid inward).  The compressive hoop stress arises as a consequence of
this solid flow ($V_r/r<0$).  Note the contrast with torsional flow,
where the compressive hoop stress is the cause of base state
segregation.

We showed above (Fig.~\ref{fig:basestatecompare}) that plane and pipe
Poiseuille are qualitatively similar in their pattern of base state
flow, but differ quantitatively.  This is evident especially in the
vertical component of the flow, which is slower in cylindrical
geometry. This can be understood as being simply related to the mass
of aggregate that is supported by a section of the wall of unit length
in the cross-flow direction.  In plane Poiseuille, the supported
material forms a rectangular column, whereas in pipe Poiseuille, the
supported material forms a shape like a slice of cake. For a pipe
radius equal to the half-width of the plane gap, the rectangular
column obviously contains more mass. Given this difference in the
vertical component, it is interesting that, for $R\lesssim0.1$, the
horizontal component of the solid velocity is similar in magnitude
between pipe and plane Poiseuille. It is also notable that in this
range of compaction lengths, where porosity bands are expected to be
prominent, pipe flow has weaker band growth relative to base state
segregation (Fig.~\ref{fig:porosityevolution}).

The time-evolution of porosity under base state segregation brings out
a problematic feature of the model: there is no physical mechanism in
the theory to balance the accumulation of liquid at the pipe wall.  It
is possible that such accumulation could occur in experiments, but
past experimental works shows that porosities are limited to
$\lesssim$25\%, even at very large strains \citep{king10}. This lack
of stabilising mechanism in the theory is an issue for all published
models of forced, laboratory deformation of partially molten
aggregates (though see \cite{takei09} for a possible solution).

Our analysis of harmonic perturbations of porosity produced results
entirely consistent with previous work on plane Poiseuille by
\cite{takei13}.  Porosity bands are expected to emerge near the pipe
wall at angles of $15$--$20^\circ$ to the vertical, if anisotropy is
at or near saturation.  As with previous analysis, the compaction
rates associated with band growth must be of the same order or larger
than those associated with base state segregation to achieve
exponential growth of infinitesimal perturbations (linear
instability). \cite{katz13} showed for plane geometry that nonlinear
interactions between base state and perturbation flow will modify both
modes, but not obscure them entirely.  We have not addressed these
interactions for pipe flow.  Moreover, we have considered only
axisymmetric, infinitesimal perturbations, which likely restrict the
behavioural space of solutions.

While comparisons with theory for torsional and plane Poiseuille flow
elucidate subtleties in the modelled dynamics, comparison with
experiments would address a more fundamental question: does outward,
base state segregation of liquid occur in synthetic, partially molten
mantle rocks subjected to forced flow through a pipe?  In experiments,
it would be necessary to force the flow with an imposed pressure
gradient, rather than with the gravitation body force.  Moreover, the
finite length of the experimental pipe would introduce complexities
not considered here.  Far from the ends of the pipe, however, we would
expect the predictions developed above to hold, if the aggregate has
an anisotropic viscosity similar to the model of
\cite{takei09a,takei09b} and \cite{takei13}.

\section{Summary and conclusions}
\label{sec:conclude}

This manuscript considered the problem of gravity-driven flow of a
partially molten aggregate through a cylindrical pipe. It presented
solutions to the equations thought to govern magma/mantle interaction,
incorporating an anisotropic viscosity tensor as a constitutive law
for the two-phase flow.  These solutions were obtained to zeroth and
first order for a linearised version of the equations, as well as to
the full, nonlinear system.

As in previous studies, anisotropic viscosity is predicted to lead to
melt segregation driven by a gradient in shear stress.  For pipe
Poiseuille geometry, this means that the liquid is expected to migrate
toward the pipe wall, causing decompaction at the outer radii of the
flow and compaction at the inner radii.  Furthermore, the
porosity-weakening of viscosity is expected to give rise to linear
instability of bands of high porosity.  Our model of anisotropic
viscosity indicates that these would take a low angle to the local
shear plane.  We have noted, however, that band growth in pipe
Poiseuille is predicted to be weaker than band growth under plane
Poiseuille.

The results presented here are consistent with previous work on
anisotropic viscosity, but extend it to pipe Poiseuille flow. This
geometry is amenable to laboratory experiments and we hope that future
work by experimentalists will evaluate the theory of anisotropic
viscosity by testing our predictions.  Ideally, a comparison with
experiments will yield insights that motivate and constrain refinement
of the model.

\paragraph{Acknowledgements} The authors thank Y.~Takei for her
comments on an early draft and acknowledge two anonymous reviews that
helped to improve the manuscript.  J.A.~was supported by a Research
Experience Placement grant from the UK Natural Environment Research
Council for Summer 2013. R.K.~is grateful for the support of the
Leverhulme Trust.

\bibliographystyle{abbrvnat} 
\bibliography{manuscript.bib}

\appendix
\section{Analytical solution for uniform anisotropy base state}
\label{sec:analytical_solution}

With both $\alpha$ and $\Theta$ constant, a suitable
transformation puts equation~\eqref{eq:basestate_gov1} into the form
of a forced, modified Bessel equation of order
$\sqrt{(C+2)/[B-D^2/(A-C+1)]}$. The solution to this equation that
satisfies the boundary condition at $r=1$ and is finite at $r=0$ is
given explicitly as
\begin{equation}
  \label{eq:besselsoln}
  V^{(0)}_r(r)=\left(\sum_{n=1}^{\infty} a_n r^n\right) - 
  \frac{I_{\sqrt{\omega_2}}(\sqrt{\omega_1}r)}
  {I_{\sqrt{\omega_2}}(\sqrt{\omega_1})}\left(\sum_{n=1}^{\infty}
    a_n\right), 
\end{equation}
where $I_{\nu}$ denotes the modified Bessel function of the first kind
of order $\nu$ \citep{baricz10} and  
\begin{align}
  a_n &= \begin{cases}
    0 & \text{for $n$ odd},\\[2mm]
    \dfrac{\omega_3}{4-\omega_2} & \text{for $n=2$}, \\[4mm]
    \dfrac{\omega_1}{n^2-\omega_2}a_{n-2} & \text{for $n$ even,
      $n>2$},
  \end{cases} \\
  \text{and}\quad \omega_1
  &=\frac{\rxi+\frac{4}{3}}{R^2(B-\frac{D^2}{A-C+1})}, \quad \omega_2
  =\frac{C+2}{B-\frac{D^2}{A-C+1}}, \quad \omega_3
  =\frac{D}{(A-C+1)B-D^2}.\label{eq30}
\end{align}
This is the solution for any constant $\alpha$ and
$\Theta$, provided $\omega_2 \neq n^2$ for $n=0$, $2$, $4$, $6$, $8$,
\ldots.

Having found the radial component of the base state velocity, the
vertical component that satisfies equation \eqref{eq:basestate_gov2}
and $V^{(0)}_z(1)=0$ is given by
\begin{equation}
  \label{eq31}
  V^{(0)}_z=\frac{r^2-1}{4(A-C+1)}+\frac{D}{A-C+1}V^{(0)}_r.
\end{equation}
We can also find a solution for $P_0(r)$ from equation~\eqref{eq:rad}.

Using the series representation of the modified Bessel function
\citep{baricz10}, we see that the boundary condition $V^{(0)}_{z,r}=0$
at $r=0$ will be satisfied if and only if $\omega_2>1$, as this is
when $I_{\sqrt{\omega_2}}$ has zero derivative at the
origin. Furthermore, even if this condition is satisfied, the solution
is not analytic at $r=0$ unless $\sqrt{\omega_2}$ happens to be an
integer. This problem is a result of the assumption that $\Theta\neq
0$ in the centre of the cylinder, which introduces a singularity at
$r=0$. If we were to use a model in which $\Theta=0$ at $r=0$, then
$V^{(0)}_r=0$ and $V^{(0)}_{z,r}=0$ at $r=0$ would follow straight
away from equations \eqref{eq:basestate_gov}.

\section{Perturbation equations and growth rate}
\label{sec:perturbation_growth}

Substituting equations~\eqref{eq:expansion} and
\eqref{eq:stresscomponents} into equations~\eqref{eq:cmp},
\eqref{eq:rad} and~\eqref{eq:zed} and then equating terms at
$O(\epsilon)$ yields
\begin{subequations}
  \begin{align}
    \nabla\cdot\boldsymbol{V}^{(1)} =&
    \frac{R^2}{r_\xi+\frac{4}{3}}\left[\frac{1}{r}
      \frac{\partial}{\partial r} \left(r\frac{\partial P_1}{\partial
          r} +\frac{rn\phi_1}{\phi_0} \frac{\partial P_0}{\partial
          r}\right) +\frac{\partial}{\partial z}\left(\frac{\partial
          P_1}{\partial z}
        +\frac{n\phi_1}{\phi_0}\right)\right], \label{eq:perturbationc1} \\
    \frac{\partial P_1}{\partial r}=&\frac{\partial}{\partial
      r}\left[A\frac{\partial V^{(1)}_z}{\partial z} +B\frac{\partial
        V^{(1)}_r}{\partial r} +C\frac{V^{(1)}_r}{r}
      -D\left(\frac{\partial V^{(1)}_z}{\partial  r}+
        \frac{\partial V^{(1)}_r}{\partial z}\right)\right] \nonumber\\
    &-\frac{\partial}{\partial r}\left[\lambda\phi_1\left(B
        \frac{\partial V^{(0)}_r}{\partial r}
        +C\frac{V^{(0)}_r}{r}-D\frac{\partial V^{(0)}_z}{\partial  r}\right)\right] \nonumber\\
    &+ \frac{1}{r}\left[(A-C)\frac{\partial V^{(1)}_z}{\partial z}
      +(B-C)\frac{\partial V^{(1)}_r}{\partial r}-2\frac{V^{(1)}_r}{r}
      -D\left(\frac{\partial V^{(1)}_z}{\partial r}+\frac{\partial
          V^{(1)}_r}{\partial z}\right)
    \right] \nonumber \\
    &-\lambda\phi_1\left((B-C)\frac{\partial V^{(0)}_r}{\partial r}
      -2\frac{V^{(0)}_r}{r}-D\frac{\partial V^{(0)}_z}{\partial r}
    \right) \nonumber \\
    &+\frac{\partial}{\partial z}\left[-E\frac{\partial
        V^{(1)}_z}{\partial z} -D\frac{\partial V^{(1)}_r}{\partial
        r}+(A-C+1)\left(\frac{\partial V^{(1)}_z}{\partial r}
        +\frac{\partial V^{(1)}_r}{\partial z} \right) \right] \nonumber \\
    &-\frac{\partial}{\partial
      z}\left[\lambda\phi_1\left(-D\frac{\partial V^{(0)}_r}{\partial
          r}
        +(A-C+1)\frac{\partial V^{(0)}_z}{\partial  r}\right) \right], 
    \label{eq:perturbationpr}\\
    \frac{\partial P_1}{\partial z} =&\left(\frac{\partial}{\partial
        r} +\frac{1}{r}\right)\left[-E\frac{\partial
        V^{(1)}_z}{\partial z} -D\frac{\partial V^{(1)}_r}{\partial
        r}+(A-C+1)\left(\frac{\partial V^{(1)}_z}{\partial r}
        +\frac{\partial V^{(1)}_r}{\partial z} \right)\right]
    \nonumber
    \\
    &-\left(\frac{\partial}{\partial r}
      +\frac{1}{r}\right)\left[\lambda\phi_1\left(-D \frac{\partial
          V^{(0)}_r}{\partial r}
        +(A-C+1)\frac{\partial V^{(0)}_z}{\partial r}\right) \right] \nonumber \\
    &+ \frac{\partial}{\partial z}\left[F\frac{\partial
        V^{(1)}_z}{\partial z} +A\frac{\partial V^{(1)}_r}{\partial
        r}+C\frac{V^{(1)}_r}{r} -E\left(\frac{\partial
          V^{(1)}_z}{\partial r}+
        \frac{\partial V^{(1)}_r}{\partial z} \right) \right]\nonumber \\
    &-\frac{\partial}{\partial z}\left[\lambda\phi_1\left(A
        \frac{\partial V^{(0)}_r}{\partial r}
        +C\frac{V^{(0)}_r}{r}-E\frac{\partial V^{(0)}_z}{\partial
          r}\right)\right].
    \label{eq:perturbationpz}
  \end{align}
\end{subequations}

When we consider the perturbation defined by
equations~\eqref{eq:porosityperturb} and \eqref{eq:othervars} in the
limit $k\rightarrow\infty$, the above equations (at leading order in
$k$) simplify to
\begin{equation}
 \begin{pmatrix}
   N_6 & N_1 & N_2 \\ N_1 & N_3 & N_4 \\ N_2 & N_4 & N_5
 \end{pmatrix}\begin{pmatrix} \tilde{P} \\ \tilde{V_r} \\ \tilde{V_z}
 \end{pmatrix} =\begin{pmatrix}W_1 \\ W_2 \\ W_3
 \end{pmatrix} \label{eq:matrix} ,
\end{equation}
with $N_1$, $N_2$, $N_3$, $N_4$, $N_5$, $W_2$, and $W_3$ as defined in
equations~\eqref{eq:growthrate_factors}, and
\begin{align}
  N_6 &= \frac{R^2}{r_\xi+\frac{4}{3}} k^2, \nonumber \\
  W_1 &= \frac{R^2}{\left(r_\xi+\frac{4}{3}\right)}\frac{
    ni}{\phi_0}\left(k_r\frac{\partial P_0}{\partial r}+k_z\right)
  \nonumber.
\end{align}
 Equation~\eqref{eq:matrix}
can be inverted to give expressions for $\tilde{P}$, $\tilde{V_r}$ and
$\tilde{V_z}$ which are valid to leading order in $k$.  In particular,
we find from the solution for $\tilde{P}$ that
\begin{align}
  \frac{1}{N_3N_5-{N_4}^2}[(N_1N_5-N_2N_4)W_2+(-N_1N_4+N_2N_3)W_3]
  &=W_1-J k^2 \tilde{P}
  \nonumber \\
  &= ik_r\tilde{V_r}+ik_z \tilde{V_z} \label{eq:growthratesolution} .
\end{align}
Finally, to obtain the growth rate stated in
equation~\eqref{eq:growthrate}, we substitute
equation~\eqref{eq:growthratesolution} into
equation~\eqref{eq:growthrate_equation}.

\section{Numerical methods for full, nonlinear solutions}
\label{sec:numerical_implementation}

The governing equations~\eqref{eq:governing_nondim} and model for
dynamic anisotropy \eqref{eq:alpha_dynamic}--\eqref{eq:Theta_dynamic}
are discretised on a regularly spaced, fully staggered Cartesian grid
in two dimensions. The elliptic system \eqref{eq:cmp}--\eqref{eq:zed}
is solved separately from the hyperbolic equation \eqref{eq:mass}.
For the latter we use a semi-implicit discretisation in time; the
flux-divergence term is discretised with a second-order Fromm upwind
scheme \citep{fromm68}. Both systems are solved using a preconditioned
Newton-Krylov method in the PETSc software framework
\citep{petsc-manual, petsc-homepage}.  The tolerance on the $L_2$
norm of the nonlinear residual is $10^{-6}$ in both cases. Further
details are provided by \cite{katz07}.

At each time-step, we first update the pressure and velocity variables
by solving the elliptic system, then we step the porosity forward in
time. We do not iterate this split solve because our tests show that
for appropriately small time-steps, the difference in the results is
negligible. Furthermore, we use the stress field from the previous
time-step to compute the anisotropy distribution applied for the
elliptic solve.  As discussed by \cite{katz13}, this avoids the
requirement of incorporating the viscosity parameters as explicit
variables in the Newton scheme; it also has an insignificant effect on
the solution.

A previous stress field is not available when computing the initial
velocity--pressure solution, hence we build up that solution using a
Picard iteration on the anisotropy parameters.  These are initialised
as uniform ($\alpha=2,\,\Theta=\pi/4$) and then updated after each
iteration of the solver.  We iterate to a solution tolerance on the
nonlinear residual of~$10^{-4}$.

\end{document}